# DenseNet for Breast Tumor Classification in Mammographic Images

Yuliana Jiménez Gaona [123*][0000-0001-7155-5546], María José Rodriguez-Alvarez[2] [0000-0001-8333-8792], Hector Espinó Morató[2][0000-0002-4089-1368], Darwin Castillo Malla[123][0000-0002-1800-1189], and Vasudevan Lakshminarayanan[34*][0000-0002-3473-1245]

[1] Departamento de Química y Ciencias Exactas, Universidad Técnica Particular de Loja, San Cayetano Alto s/n CP1101608, Loja, Ecuador
[2] Instituto de Instrumentación para la Imagen Molecular I3M, Universitat Politécnica de Valencia, E-46022 Valencia, Spain
[3] Theoretical and Experimental Epistemology Lab, School of Optometry and Vision Science,
[4] Department of Systems Design Engineering, Physics, and Electrical and Computer Engineering, University of Waterloo, Waterloo, ON N2L3G1, Canada
ydjimenez@utpl.edu.ec, mjrodri@i3m.upv.es, hespinos@i3m.upv.es, dpcastillo@utpl.edu.ec,vengulak@uwaterloo.ca

**Abstract.** Breast cancer is the most common invasive cancer in women, and the second main cause of death. Breast cancer screening is an efficient method to detect indeterminate breast lesions early. The common approaches of screening for women are tomosynthesis and mammography images. However, the traditional manual diagnosis requires an intense workload by pathologists, who are prone to diagnostic errors. Thus, the aim of this study is to build a deep convolutional neural network method for automatic detection, segmentation, and classification of breast lesions in mammography images. Based on deep learning the Mask-CNN (RoIAlign) method was developed to features selection and extraction; and the classification was carried out by DenseNet architecture. Finally, the precision and accuracy of the model is evaluated by cross validation matrix and AUC curve. To summarize, the findings of this study may provide a helpful to improve the diagnosis and efficiency in the automatic tumor localization through the medical image classification.

**Keywords:** breast tumor classification, convolutional neural network, mammography, RoI Align, DenseNet, deep learning.

## 1 Introduction

Breast cancer screening is an efficient method to detect indeterminate breast lesions early [1-6]. Clinically, the best approaches of screening for women are ultrasound [7] and mammography [8-9] images. After the screening procedure if there are suspicious lesions, the analysis could be combined with biopsies [10], histopathological images [11-13] and magnetic resonance imaging (MRI) [14].

The ultrasound allows obtaining high quality images, without the need for ionizing radiation, and enables detection of very small lesions, even masses and microcalcifications (MC). However, mammography (x-rays) is currently the most used imaging method to detect breast cancer early in both patients, in both symptomatic and asymptomatic patients [2], reducing unnecessary biopsies. It is recommended by WHO, as the standard imaging for early diagnosis.

Specialists can interpret the breast images with the latest breast imaging reporting and data system (BI-RADS) version, proposed by the American College of Radiology [15-17]. Nevertheless, the traditional manual diagnosis is time consuming and prone to diagnostic errors [18,19]. Multidimensional digital images from physiological structures can be processed and manipulated to help visualize hidden diagnostic features [20]. For this purpose, techniques based of Deep Learning (DL) and Machine Learning (ML) [20-25], are used to improve the diagnosis and efficiency in the location and tumor processes monitoring through the automatic medical image classification. Convolutional neural networks (CNN), an extensively used DL methodology has been used to analyze medical images [27-33]. A recent paper reviews various aspect of applications of convolutional neural networks to breast cancer detection and automated diagnosis, Jiménez et al. [27] provides a critical review of the literature on DL applications in breast tumor diagnosis using ultrasound and mammography images. The main findings in the classification process revealed that new DL-CAD methods are useful and effective screening tools for breast cancer, reducing the need for manual feature extraction.

We have developed a novel CNN for automatic extraction, selection and classification of breast lesions from mammography images, in this paper we present the network and discuss the results obtained from this network.

## 2 Materials and methodology

The process for classifying the breast tumors is illustrated in Fig. 1 and the steps are as follows: (1) Breast Dataset acquisition and Preprocessing. (2) RoI image segmentation, feature selection and extraction using a Mask R-CNN with RoIAlign technique. (3) Breast tumor classification using DenseNet architecture. (4) The evaluation performance metrics. The Mask R CNN and RoIAlign are discussed below.

### 2.1 Dataset

We used a public Breast Cancer Digital Repository (BCDR, https://bcdr.eu/) database, for training and evaluation of the CNN. The BCDR-F03 [34] mammography dataset has 736 biopsy-proven lesions from 1734 patients. Each case includes clinical data for each patient, and both Cranio-caudal (CC) and Medio-lateral oblique (MLO) view mammograms, along with the coordinates of the lesion contours, and a binary class dataset composed of benign and malignant findings.

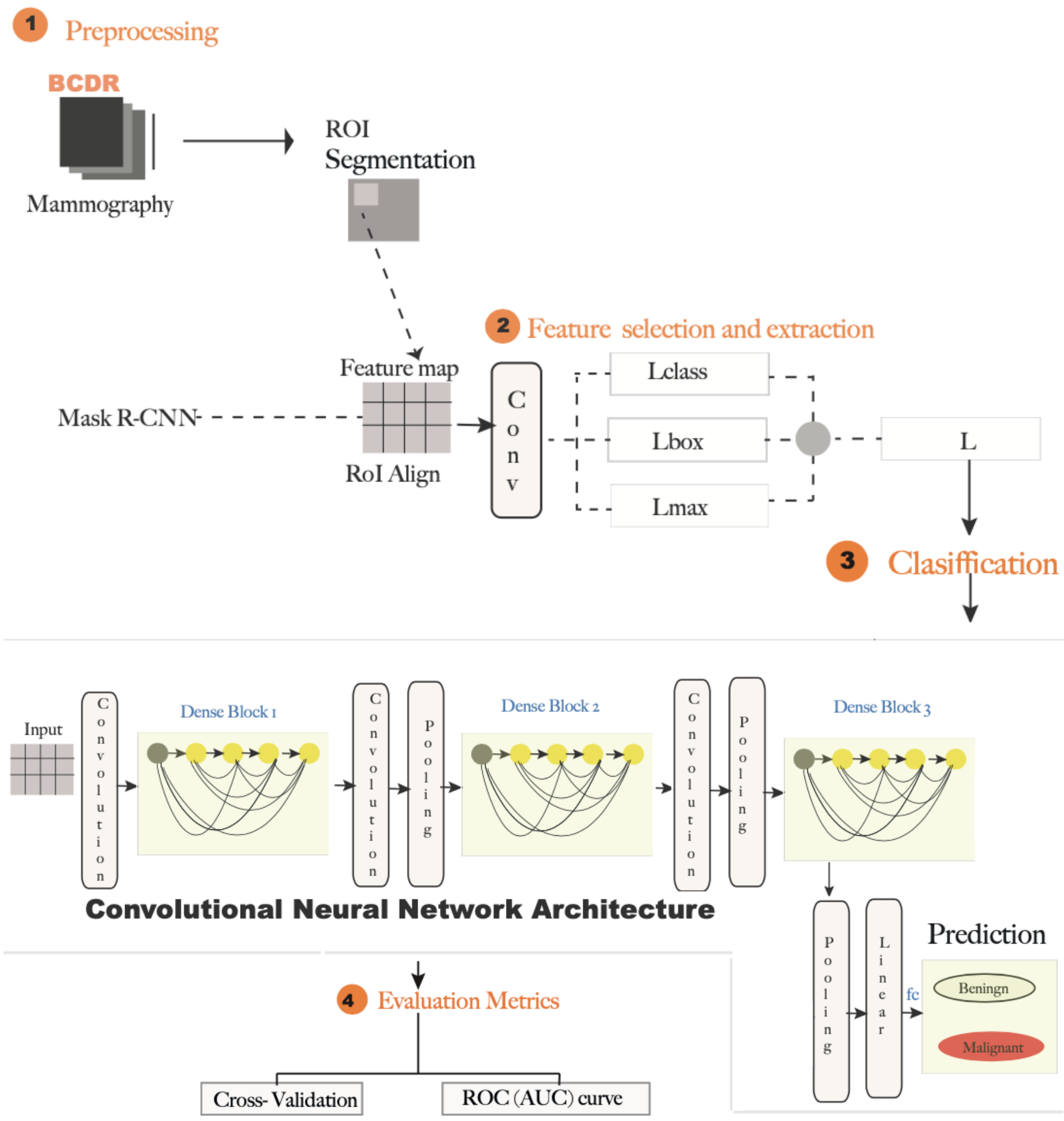

**Fig. 1.** Illustration of the Breast-Dense workflow. Deep Convolutional neural network (DCNN) for Breast Cancer tumor classification.

## 2.2 Segmentation and Feature Extraction

Preprocessing to remove noise and artifacts in breast cancer in particular consists of delineation of tumors from the background, breast border extraction and pectoral muscle suppression [24]. Then, the images are segmented for ROI extraction, the regions being the possible tumors. This operation provides us the coordinates to target and crop the bounding box of the lesions automatically.

Once the ROI is detected and cropped, we extract the features of the tumor contour by a Mask R-CNN [35] network trained using RoI alignment (RoI Align) technique. This technique is based on bilinear interpolation to smoothly crop a patch from a full-image feature maps based on a region proposal network (RPN), and then resize the cropped patch to a desired spatial size using a loss function. This has shown to outperform the use of ROI pooling [28].

According to Fig. 2, the four sampling points in each bin dashed grid represents the RoIAlign method. It computes the value of each sampling point by bilinear interpolation from the nearby grid points on the feature map. The RoIPooling uses max pooling to convert features in the projected region of the image of any size, (x1) x (y1), into a small fixed window, [x1] x [y1]. The input region is then divided into [x1] and [y1] grids, giving approximately every sub-window of size ([x1]/x1) ([y1]/y1). Then max-pooling us then applied to every grid.

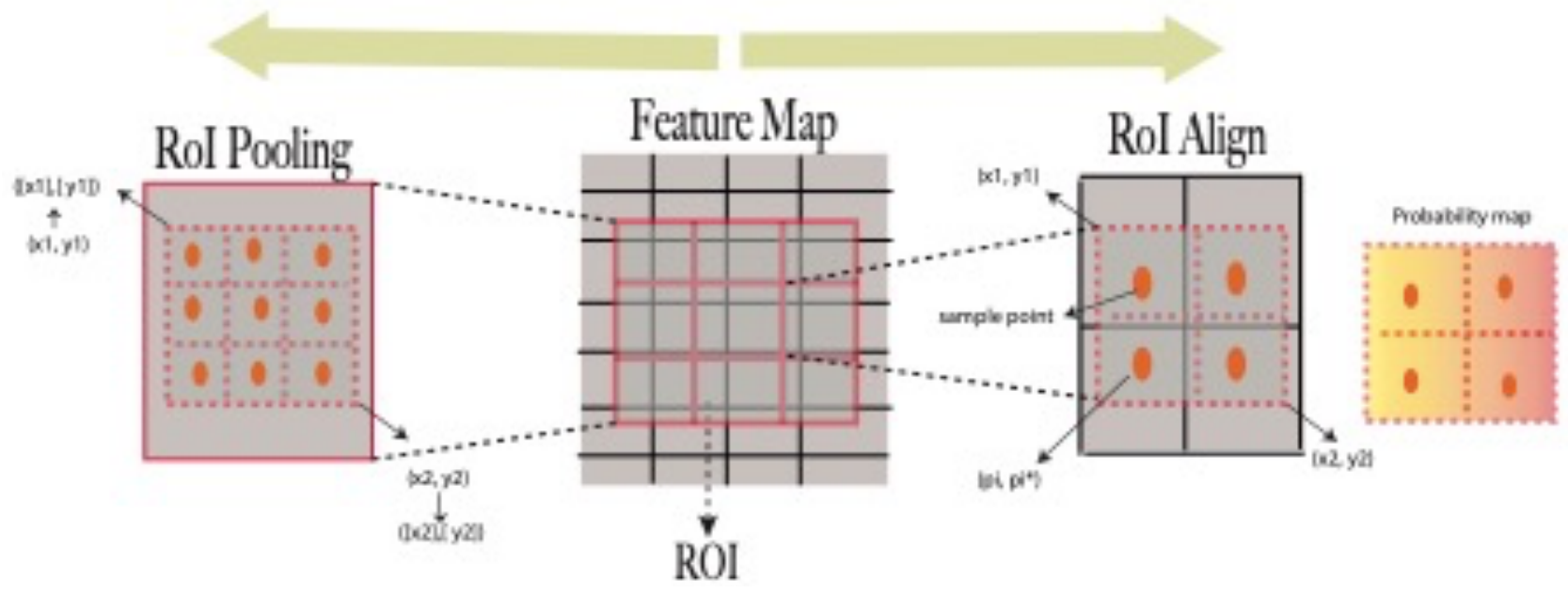

**Fig. 2.** RoIPooling and RoIAlign illustration, from the feature map.

During the Mask R-CNN training, the values of the loss function (L, $L_{class}$, $L_{box}$, $L_{mask}$) is minimized,

$$L = L_{class} + L_{box} + L_{mask} \tag{1}$$

where L represents the loss function, $L_{class}$ is the classification loss, $L_{box}$ is the bounding-box loss regression and $L_{mask}$ is the average binary cross-entropy loss mask prediction. Also, $L_{class}$ + $L_{box}$, and $L_{class}$ are defined by equations (2) and (3)

$$L_{class+}L_{box} = \frac{1}{N_{cls}}\sum_i L_{cls}\left(p_i, p_i^*\right) + \frac{1}{N_{box}}\sum_i p_i^* L_1^{smooth}(t_i - t_i^*) \tag{2}$$

$$L_{cls}\left(\left\{p_i, p_i^*\right\}\right) = -p_i^* \log p_i^* - (1 - p_i^*)\log(1 - p_i^*) \tag{3}$$

where, *smooth* in equation (2) is given by:

$$Smooth_{L1}(x) = \left\{ \begin{array}{ll} 0.5x^2 & if\ |x| < 1 \\ |x| - 0.5 & otherwise, \end{array} \right\} \tag{4}$$

and, the $L_{mask}$, is:

$$L_{mask} = -\frac{1}{m^2}\sum_{1 \le i,j \le m}\left[y_{ij}\log \hat{y}_{ij}^k + \left(1 - y_{ij}\right)\log(1 - \hat{y}_{ij}^k)\right] \tag{5}$$

The different variables are interpreted in table 1.

**Table 1.** Definition of different variables of the equation (1), (2), (3).

| Symbol | Explanation |
|---|---|
| $P_i$ | Predicted probability of anchors *i*, being an object. |
| $p_i^*$ | Ground truth label (binary) of whether anchor *i* is an object. |
| $t_i$ | Predicted four parameterized coordinates. |
| $t_i^*$ | Ground truth coordinates. |
| $N_{cls}$ | Normalization term, set to be mini batch size ~ 256 |
| $N_{box}$ | Normalization term, set the number of anchor locations ~ 2400 |
| $y_{ij}$ | Is the label of a cell (*i, j*) in the mask for the region of size m x m. |
| $\hat{y}_{ij}^{k}$ | Is the predicted value of the same cell in the mask learned for the ground truth class k. |

## 2.3 Classification: DenseNet Architecture

We used the DenseNet CNN architecture for evaluation and analysis of our dataset, as in Table 2. DenseNet presents, several advantages over other pretraining CNN methods: effectively solve the vanishing-gradient problem, reduce the number of parameters, the feature reuse, and strengthen feature propagation [18], (see Fig. 3).

In addition, the CNN is a sequence of feedforward layers implementing convolutional filters and pooling layers. After the last pooling layer, the CNN adopts several fully connected layers that work on converting the 2D feature maps of the previous layers into 1D vector for classification [22]. This can be represented as:

$$G(X) = g_N(g_N - 1(...(g_1(x))) \tag{6}$$

where $N$ represents number of hidden layers, $X$ is the input signal and $g_N$ denotes the corresponding function to the layer $N$. A basic CNN model has a convolutional layer which consists of a function $g$, with multiple convolutional kernels $(h_1, \ldots h_{k-1}, h_k)$. Every $h_k$ denotes a linear function in $k$th kernel, represented as follows (7):

$$h_k(x,y) = \sum_{s=-m}^{m} \sum_{t=-n}^{n} \sum_{v=-d}^{w} V_k(s,t,v) X(x-s, y-t, z-v) \tag{7}$$

where $(x, y, z)$ represents pixel position of input $X$, $m$ represents height, $n$ denotes width, $w$ is depth of the filter, and $V_k$ represents weight of $k$th kernel. A schematic flowchart of CNN is shown in Fig. 3 and Table 2.

**Table 2.** DenseNet architecture. The growth rate for all the networks is k=32. Note that each "conv" layer shown in the table corresponds the sequence BN-ReLU-Conv

| Layers | Output Size | DenseNet-121 | DenseNet-169 | DenseNet-201 | DenseNet-264 |
|---|---|---|---|---|---|
| Convolution | 112 x 112 | 7 x 7 conv, stride 2 | | | |
| Pooling | 56 x 56 | 3 x 3 max pool, stride 2 | | | |
| Dense Block 1 | 56 x 56 | $\begin{vmatrix} 1\ x\ 1\ conv \\ 3\ x\ 3\ conv \end{vmatrix}\ x\ 6$ | $\begin{vmatrix} 1\ x\ 1\ conv \\ 3\ x\ 3\ conv \end{vmatrix}\ x\ 6$ | $\begin{vmatrix} 1\ x\ 1\ conv \\ 3\ x\ 3\ conv \end{vmatrix}\ x\ 6$ | $\begin{vmatrix} 1\ x\ 1\ conv \\ 3\ x\ 3\ conv \end{vmatrix}\ x\ 6$ |
| Transition Layer 1 | 56 x 56 | 1x1 conv | | | |
| | 28 x 28 | 2 x 2 average pool, stride 2 | | | |
| Dense Block 2 | 28 x 28 | $\begin{vmatrix} 1\ x\ 1\ conv \\ 3\ x\ 3\ conv \end{vmatrix}\ x\ 12$ | $\begin{vmatrix} 1\ x\ 1\ conv \\ 3\ x\ 3\ conv \end{vmatrix}\ x\ 12$ | $\begin{vmatrix} 1\ x\ 1\ conv \\ 3\ x\ 3\ conv \end{vmatrix}\ x\ 12$ | $\begin{vmatrix} 1\ x\ 1\ conv \\ 3\ x\ 3\ conv \end{vmatrix}\ x\ 12$ |
| Transition Layer 2 | 28 x 28 | 1x1 conv | | | |
| | 14 x 14 | 2 x 2 average pool , stride 2 | | | |
| Dense Block 3 | 14 x 14 | $\begin{vmatrix} 1\ x\ 1\ conv \\ 3\ x\ 3\ conv \end{vmatrix}\ x\ 24$ | $\begin{vmatrix} 1\ x\ 1\ conv \\ 3\ x\ 3\ conv \end{vmatrix}\ x\ 32$ | $\begin{vmatrix} 1\ x\ 1\ conv \\ 3\ x\ 3\ conv \end{vmatrix}\ x\ 48$ | $\begin{vmatrix} 1\ x\ 1\ conv \\ 3\ x\ 3\ conv \end{vmatrix}\ x\ 64$ |
| Transition Layer 3 | 14 x 14 | 1x1 conv | | | |
| | 7 x 7 | 2 x 2 average pool , stride 2 | | | |
| Dense Block 4 | 7 x 7 | $\begin{vmatrix} 1\ x\ 1\ conv \\ 3\ x\ 3\ conv \end{vmatrix}\ x\ 16$ | $\begin{vmatrix} 1\ x\ 1\ conv \\ 3\ x\ 3\ conv \end{vmatrix}\ x\ 32$ | $\begin{vmatrix} 1\ x\ 1\ conv \\ 3\ x\ 3\ conv \end{vmatrix}\ x\ 32$ | $\begin{vmatrix} 1\ x\ 1\ conv \\ 3\ x\ 3\ conv \end{vmatrix}\ x\ 48$ |
| Classsification Layer | 1 x 1 | 7x7 global average pool | | | |
| | | 1000D fully connected, softmax | | | |

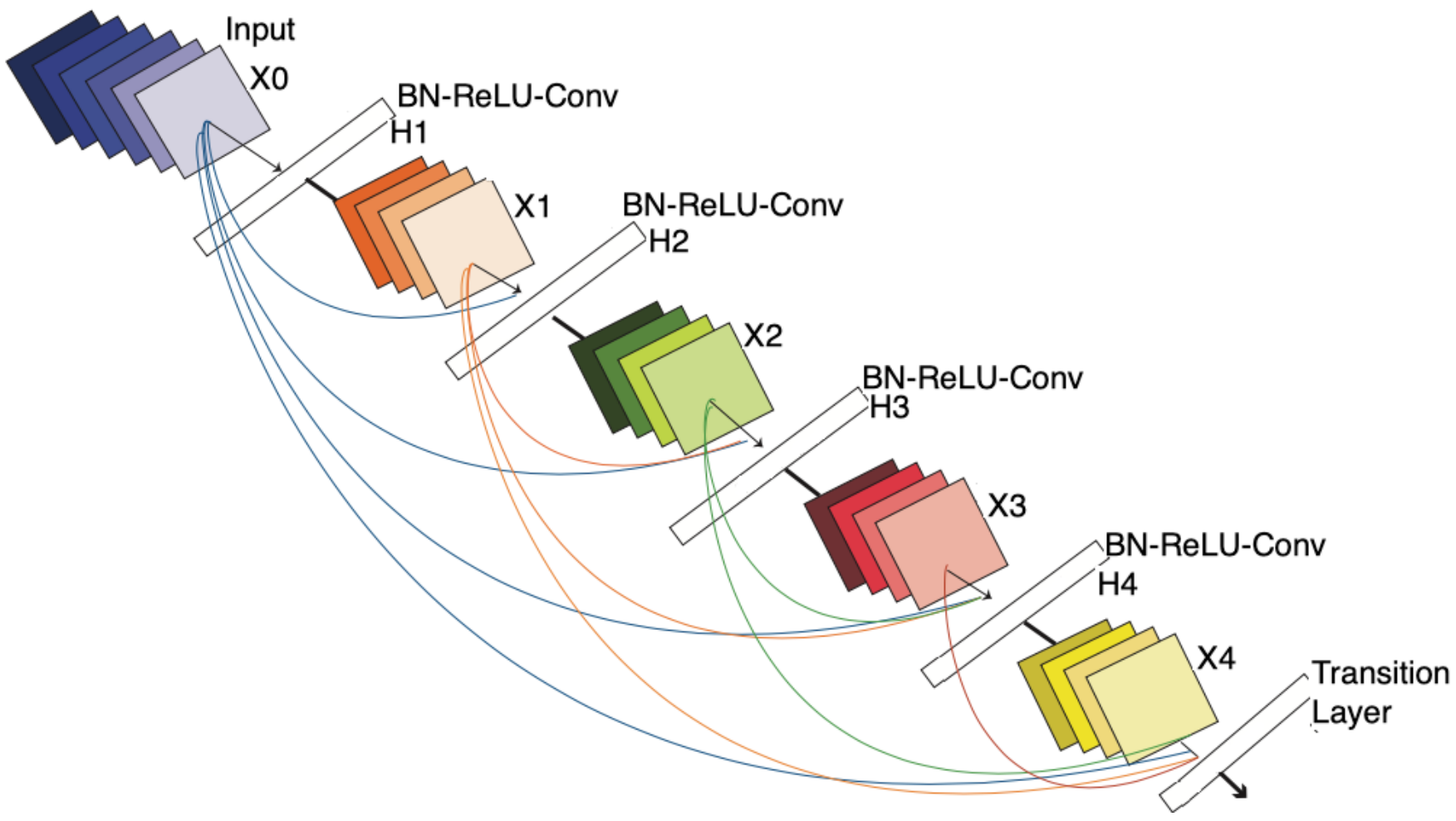

**Fig. 3.** A 5-layer dense block with a growth rate of k = 4. Each layer takes all preceding feature-maps as input. These layers reducing the amount of computation and improve the robustness.

### 2.4 Evaluation Metrics

Different quantitative metrics are used to evaluate the classifier performance of a DL system [33]. These include *Accuracy* (Acc), *Sensitivity* (Sen), *Specificity* (Spe), *Area Under the Curve* (AUC), *Precision, F1 score*.

The performance of the trained Mask R-CNN model was quantitatively evaluated by mean average precision (MAP) as the accuracy of lesion detection/segmentation on the validation set (8):

$$mAP = \frac{A \cap B}{A \cup B} = \frac{1}{N_T} \sum_{i}^{N_t} \left( \frac{N_i^{DR}}{N_i^{D}} \right) \tag{8}$$

where A is the model segmentation result, and B is the contour tumor delineated by the experienced radiologist. $N_T$ is the number of images; $N_i^{DR}$ is the overlapped area between the model detected lesion and the true clinical lesion regions; $N_i^{D}$ is the size of the true clinical lesion.

## 3 Results

In this study, the 344 cases in the image database (178 benign and 166 malignant) were split into 80% as the training set and 20% as the validation set.

The performance of the trained Mask R-CNN model achieved a MAP value of 0.75 for the automatic lesion delineation in validation dataset.

### 3.1 Breast DenseNet

The Table 3 summarizes the results of the Breast DenseNet model using the BCDR dataset, and their comparison performance evaluation with different pre-trained models in terms of the acc, sensitivity, specificity and AUC.

**Table 3.** Results summary of pre-trained DL models in mammograms.

| Reference | Method | Database | Sensitivity (%) | Specificity (%) | AUC | Acc (%) |
|---|---|---|---|---|---|---|
| Al-Masni et al.[36] | YOLO5- Fold cross validation. | DDSM | 100 | 94 | 96.5 | 97 |
| Ragab DA et al.[2] | CNN + Linear SVM | DDSM | 77 | 84 | 88 | 80.5 |
| Duggento A. et al. [21] | CNN | CBIS-DDSM | 84.4 | 62.4 | 77 | 71 |
| H. Choudgrad et al. [23] | CNN | DDSM | - | - | 98 | 97.4 |
| Debelee et al [37] | | MIAS<br>DDSM | 96.26<br>99.48 | 100<br>98.16 | -<br>- | 97.46<br>99 |
| Ahmed et al [38] | | Inbreast | 80 | - | 78 | 80.10 |
| Jimenez et al. | DenseNet | BCDR | 99 | 94 | 97 | 97.7 |

## 4 Discussion

Alkhaleefah, et al., [39] used transfer learning technique to classify benign and malignant breast cancer by CNN networks such as AlexNet, visual geometry group (VGG), GoogLeNet, and residual network (ResNet) on breast cancer datasets. However, these networks have been trained on large datasets such as ImageNet, which do not contain labeled images related to breast cancers, which lead to poor performance.

Thus, according to [33] the most utilized databases for mammography images are MIAS, DDSM, BCDR, and Inbreast. In this study, we have selected the BCDR database because it contains cases of 1734 patients with mammography and ultrasound images, clinical history, lesion segmentation, patient cases are BIRADS classified and annotated by specialized radiologists.

Also, we developed a Breast-DenseNet Deep learning system, to detect the locations of potential masses on mammograms and classify them into benign or malignant. We did not require filtering and noise elimination before segmentation and feature extraction to improve the accuracy [36]. The ROI regions were automatically delineated and the feature extraction tumor was done via YOLOv3 based on Mask RCNN.

AlMasni [36], specified two important issues faced by the YOLO approach in the clinical mammographic field. First, it could reveal the breast masses, which existed over the pectoral muscle and second, the proposed methodology successfully identified breast masses in the dense tissues. Further, the traditional studies used support vector machine (SVM) [2,40] a method of machine learning, in detection and classification. Those methods needed to extract features from ROI and then the features were given to SVM classifier through SVM detection of benign and malignant lesions in breast ultrasound images using texture morphological and fractal features. However, in this work it was not necessary because we develop a Breast-Dense CNN for automatic detection, segmentation, and classification of breast lesions.

## 5 Conclusions

We conclude that DL brings an apparent improvement compared to other approaches. The Breast-Dense strategy proposed improves the state-of-the-art accuracy classification on the BCDR dataset. The YOLO + DenseNet model trained on the dataset, achieved the best accuracy rate overall, and was used to develop a tumor lesion classification tool.

Breast-DenseNet provided highly accurate diagnoses when classifying benign from malignant tumors. Therefore, its predictor could be used as a second opinion to assist the radiologist diagnoses. Our future work includes deeper architectures as well as ultrasound, histopathology and PET images to deal with problems caused by mammograms of highly dense breasts. It will be helpful to include others imaging techniques, in combination with mammography during the learning process, to help to model a robust breast mass predictor. In conclusion, Table 3 demonstrated that Breast DenseNet achieved better results compared to other state-of-the-art methods, which classified the same public dataset. For instance, we achieved 97.7% accuracy and 97% AUC on the BCDR database.

**Acknowledges**

VL would like to thank the natural sciences and engineering research council of Canada (NSERC) for a discovery grant. Y.J.G. and D.C.M. acknowledges the research support of Universidad Técnica Particular de Loja through the project PROY_INV_QUI_2020_2784.